\nonumber
\documentclass[final,3p,times]{elsarticle}
\usepackage{amssymb}
\usepackage{amsmath}
\usepackage{amsthm}
\usepackage{hyperref}
\usepackage{lineno}
\usepackage{graphicx}
\usepackage{multirow}
\usepackage{tikz}
\usetikzlibrary{positioning}
\usetikzlibrary{fit,arrows, decorations.markings}
\tikzstyle{vecArrow} = [thick, decoration={markings,mark=at position
	1 with {\arrow[semithick]{open triangle 60}}},
double distance=1.4pt, shorten >= 5.5pt,
preaction = {decorate},
postaction = {draw,line width=1.4pt, white,shorten >= 4.5pt}]
\tikzstyle{sbox} = [draw=black, fill=white!15, very thick,text width=5cm,
rectangle, rounded corners, inner sep=3pt, inner ysep=6pt,align=center]
\usepackage{tabularx}
\usepackage{xspace}
\biboptions{sort&compress}
\newcounter{bla}

\journal{Computer Physics Communications}

\newcommand{\sss}[1]{\scriptscriptstyle{#1}}
\newcommand{\ds}{\displaystyle}

\newcommand{\SANC}{\texttt{SANC}\xspace}
\newcommand{\MCSANC}{\texttt{MCSANC}\xspace}

\newcommand{\ReneSANCe}{\texttt{ReneSANCe}\xspace}
\begin{document}
\begin{frontmatter}
\title{Hadron-hadron collision mode in {\tt ReneSANCe-v1.3.0}}

\author[a]{Serge~Bondarenko\corref{author}}
\ead{bondarenko@jinr.ru}
\author[b,c]{Yahor~Dydyshka\corref{author}}
\ead{yahord@gmail.com}
\author[b]{Lidia~Kalinovskaya\corref{author}}
\ead{lidia.kalinovskaya@cern.ch}
\author[b]{Renat~Sadykov\corref{author}}
\ead{sadykov@cern.ch}
\author[b,c]{Vitaly~Yermolchyk\corref{author}}
\ead{Vitaly.Yermolchyk@jinr.ru}

\cortext[author]{Corresponding authors.}
\address[a]{Bogoliubov Laboratory of Theoretical Physics, JINR, Joliot-Curie 6, RU-141980 Dubna, Russia}
\address[b]{Dzhelepov Laboratory for Nuclear Problems, JINR, Joliot-Curie 6, RU-141980 Dubna, Russia}
\address[c]{Institute for Nuclear Problems, Belarusian State University, Bobruiskaya 11, 220006 Minsk, Belarus}

\begin{abstract}

We report an implementation of the hadron-hadron ($pp$ and $p\bar{p}$) collision mode to the Monte Carlo event generator {\tt ReneSANCe} --- the code that was previously developed for $e^{+}e^{-}$
collisions. The described extension of ReneSANCe currently contains neutral and charged current Drell-Yan prosesses $pp[p\bar{p}] \to ZX \to \ell^+\ell^-X$, $pp[p\bar{p}] \to W^+X \to \ell^+\nu_\ell X$ and $pp[p\bar{p}] \to W^-X \to \ell^-\bar{\nu}_\ell X$.
We take into account complete one-loop electroweak (EW) and one-loop QCD corrections to these processes. The calculation is based on the SANC 
(Support for Analytic and Numeric Calculations for experiments at colliders) modules. The generator is constructed in such a way that new processes can be easily added. 
The paper contains a theoretical description of the SANC approach, numerical validations and a manual.

\end{abstract}

\begin{keyword}

Perturbation theory; NLO calculations; Standard Model; Electroweak interaction; QED; QCD; Monte Carlo simulation

\end{keyword}

\end{frontmatter}

{\bf PROGRAM SUMMARY}

\begin{small}
\noindent
{\em Program Title: ReneSANCe-v1.3.0}\\
{\em Licensing provisions: GPLv3}\\
{\em Programming language: Fortran, C, C++}\\
{\em Supplementary material: Looptools~[1], FOAM~[2]}\\
{\em Nature of problem: Theoretical calculations at next-to-leading order in
perturbation theory allow to compute higher precision amplitudes for Standard
Model processes and decays, provided proper treatments of UV divergences and IR
singularities are performed}\\
{\em Solution method: Numerical integration of the precomputed differential expressions for
cross sections of certain processes implemented as SANC modules~[3,4]}\\
{\em Restrictions: the list of processes is limited to $pp[p\bar{p}] \to ZX \to \ell^+\ell^-X$, $pp[\bar{p}] \to W^+X \to \ell^+\nu_\ell X$ and $pp[\bar{p}] \to W^-X \to \ell^+\bar{\nu}_\ell X$}\\

\end{small}

\section{Introduction}
\label{sec:introduction}
The construction of high-precision theoretical predictions and their comparison with experimental data play a crucial role in solving the problem of applicability of the Standard Model and justifying its structure on the basis of fundamental principles. Implementation of the results of theoretical calculations in Monte Carlo event generators is an important step of theoretical support for high-precision experimental verification of the Standard Model carried out in current and future experiments. 

Monte Carlo event generators are used to allow for detector effects in experimental data and obtain predictions with which these experimental data will be compared. In addition, they can be used to obtain histograms of complex observables and pseudo-observables without rerunning and rewriting the code. One can simply take the generated events and analyze them using programs such as {\tt RIVET}~\cite{Buckley:2010ar} or {\tt ROOT}~\cite{Brun:1997pa}, in contrast to integrators where histograms are often hardcoded.

It is very important to comprehensively take into account the effects of higher-orders corrections due to strong, electromagnetic and weak interactions.

This paper describes the extension of the Monte Carlo (MC) event generator {\tt ReneSANCe}~\cite{Sadykov:2020any} previously developed for $e^+e^-$ processes to simulate processes at hadron-hadron colliders with allowance for electroweak (EW) and QCD corrections with the next-to-leading order (NLO) accuracy and also higher-order EW corrections through $\Delta \rho$ parameter. The described extension contains at present the inclusive charged and neutral current (NC and CC) Drell-Yan (DY) processes at $pp$ and $p\bar{p}$ colliders: 
\begin{eqnarray}
& pp[p\bar{p}] \to Z + X \to \ell^+\ell^- + X,\\ 
&pp[p\bar{p}] \to W^+ + X \to \ell^+\nu_\ell + X\\ & pp[p\bar{p}] \to W^- + X \to \ell^-\bar{\nu}_\ell + X,
\end{eqnarray}
with $\ell= e,\mu$.

The Drell-Yan specific code in {\tt ReneSANCe} originates from the earlier \SANC MC integrator and generator \cite{Arbuzov:2005dd,Arbuzov:2007db,Arbuzov:2007kp,Richardson:2010gz}. Since then, it has been substantially improved and extended. These tools and the later MC integrator \MCSANC \cite{Bondarenko:2013nu, Arbuzov:2015yja} for processes at pp colliders share common \SANC Fortran modules \cite{ANDONOV2006481}. The results of these codes are thoroughly verified in the systematic comparison of public Monte Carlo codes, which describe the Drell-Yan processes at hadron colliders \cite{Buttar:2006zd, TeV4LHC-Top:2007fwh,Buttar:2008jx,Arbuzov:2012dx,Antropov:2017bed,Alioli:2016fum}. Full support of $p\bar{p}$ colliders is added in \SANC codes for the first time.

Currently, multiple independent MC event generators are available that calculate NLO EW corrections to DY processes, e.g. {\tt KKMChh}~\cite{Jadach:1999vf,Jadach:2016zsp}, {\tt WGRAD/ZGRAD}~\cite{Baur:1998kt,Baur:2001ze,Baur:2004ig}, {\tt HORACE}~\cite{CarloniCalame:2003ux,CarloniCalame:2005vc,CarloniCalame:2006zq,CarloniCalame:2007cd}, {\tt WINHAC/ZINHAC}~\cite{Placzek:2003zg,Placzek:2013moa}, {\tt POWHEG-EW}~\cite{Bernaciak:2012hj,Barze:2012tt}, {\tt RADY}~\cite{Dittmaier:2001ay,Dittmaier:2009cr}, {\tt LePaProGen}~\cite{Dydyshka:2017fie}.

The paper is organized as follows. Section~\ref{sec:physics} contains a brief description
of one-loop EW and QCD calculations in the {\tt SANC} system. The structure of the code is described in section~\ref{sec:code}. The benchmarks against {\tt MCSANC} integrator are presented in section~\ref{sec:validation}. A summary is given in section~\ref{sec:summary}.

\section{NLO EW and QCD corrections in {\tt SANC} framework}
\label{sec:physics}
The calculations are organized in a way that allows one to control the consistency of the result. All analytical calculations at the one-loop precision level are realized in the $R_\xi$ gauge with three gauge parameters: $\xi_{\mathsf{A}}$, $\xi_{\mathsf{Z}}$, and $\xi\equiv\xi_{\mathsf{W}}$. To parameterize ultraviolet divergences, dimensional regularization is used. Loop integrals are expressed in terms of the standard scalar Passarino-Veltman functions: $ A_0, \, B_0, \, C_0, \, D_0$ \cite{PASSARINO1979151}. These features make it possible to carry out several important checks at the level of analytical expressions, e.g., checking the gauge invariance by eliminating the dependence on the gauge parameter, checking cancellation of ultraviolet poles, as well as checking various symmetry properties and the Ward identities.

The cross section of the process $e^+e^- \to X$ at the one-loop EW level can be divided into four parts:
\begin{eqnarray}
\sigma^{\text{1-loop}} = \sigma^{\mathrm{Born}} + \sigma^{\mathrm{virt}}(\lambda) + \sigma^{\mathrm{soft}}(\lambda, \omega)
+ \sigma^{\mathrm{hard}}(\omega),
\label{loopxsec}
\end{eqnarray}
where $\sigma^{\mathrm{Born}}$ is the Born level cross section, $\sigma^{\mathrm{virt}}$ is the contribution of virtual (loop) corrections, $\sigma^{\mathrm{soft}}$ corresponds to the soft photon
emission, and $\sigma^{\mathrm{hard}}$ is the hard photon emission part (with energy $E_{\gamma} > \omega$). Auxiliary parameters $\lambda$ (fictitious "photon mass" which regularizes infrared divergences) and
$\omega$ (photon energy which separates the regions of the phase space associated with the soft and hard emissions) cancel out after summation.

Calculations were carried out in the {\tt SANC} framework \cite{ANDONOV2006481} using {\tt FORM} \cite{Vermaseren:2000nd}. For optimization we intensively used factorization \cite{Kuipers:2012rf} introduced in {\tt FORM} 4  and elimination
of common subexpressions with temporary variables \cite{Kuipers:2013pba}. The results of calculations were automatically transformed into software modules in the {\tt FORTRAN} 77 language with the standard {\tt SANC} interface.
Evaluation of loop integrals in the modules is performed by {\tt Looptools} \cite{Hahn:1998yk} and {\tt SANClib} \cite{Bardin:2009zz,Bardin:2009ix} packages.
These modules incorporating physical calculations can be used in any code that understands the {\tt SANC} interface \cite{Andonov:2008ga}.

\section{{\tt ReneSANCe} structure}
\label{sec:code}
\subsection{Configuration}

Configuration files support JSON-like syntax and contain general steering parameters for a run, {\tt FOAM}  parameters, kinematic cuts and parameters of the Standard Model.
The configuration can be split into several files and included in the index.conf file by \textit{.include $<$relative or absolute path to file$>$} macro.

\subsubsection{Process parameters}
\begin{description}
	\item [pid] [{\tt integer}] = defines a partonic process to calculate (\texttt{integer})\\
		201: $hh \to e^+e^- X$\\
		202: $hh \to \mu^+\mu^- X$\\
		203: $hh \to \tau^+\tau^- X$\\
		211: $hh \to e^-\bar{\nu}_e X$\\
		212: $hh \to \mu^-\bar{\nu}_\mu X$\\
		213: $hh \to \tau^-\bar{\nu}_\tau X$\\
		221: $hh \to \nu_e e^+ X$\\
		222: $hh \to \nu_\mu \mu^+ X$\\
		223: $hh \to \nu_\tau \tau^+ X$
	\item [ibeams] [{\tt integer}] = 1/2 corresponds to $pp$/$p\bar{p}$ colliding beams,
	\item [pdfname] [{\tt string}] sets the name of the PDF set connected via LHAPDF,
	\item [pdfmember] [{\tt integer}] sets a member of the PDF set,
	\item [ecm] [{\tt double}] sets collider energy in the center-of-momentum frame,
	\item [costhcut] [{\tt double}] cut on $|\cos\theta|$ for both final state particles,
	\item [ome] [{\tt double}] is the parameter separating contributions from soft and hard photon bremsstrahlung ($\omega=ome \sqrt{s}/2$).
\end{description}
	Flags controlling components of the NLO EW computations:
	\begin{description}
		\item [iqed] [{\tt integer}] = \hfill \\
		0: disables QED corrections\\
		1: with full QED corrections\\
		2: only initial state QED radiation (ISR)\\
		3: initial-final QED radiation interference term (IFI) \\
		4: only final state QED radiation (FSR)\\
		5: sum of initial and final state radiation contributions [IFI+FSR]\\
		6: sum of initial state and initial-final QED interference terms [ISR+IFI]
		\item [iew] [{\tt integer}] = 0/1 corresponds to disabled/enabled weak corrections,
		\item [iborn] [{\tt integer}] = 0 or 1 selects respectively LO or NLO level of calculations,
		\item [ifgg] [{\tt integer}] = choice of calculations for photonic vacuum polarization \(\mathcal{F}_{\gamma\gamma}\)\\
		-1: 0 \\
		0: 1 \\
		1: \(1+\mathcal{F}_{\gamma\gamma}(\mathrm{NLO})\) \\
		2: \(1/[1 - \mathcal{F}_{\gamma\gamma}(\mathrm{NLO})]\)
		\item [irun] [{\tt integer}] = 0/1 corresponds to the fixed/running gauge boson width,
		\item [gfscheme] [{\tt integer}] = flag selects the electroweak scheme in which the calculation is performed\\
		0: $\alpha(0)$-scheme\\
		1: $G_{\mu}$-scheme\\
		2: $\alpha(M_Z)$-scheme
	\end{description}

\subsubsection{FOAM parameters}
By default, optimal {\tt FOAM} parameters are used, but a user can tune some parameters manually:
\begin{description}
	\item [Nev\{Born,Virt,Hard,Brdq,Ibaa,Ibaq,Ibs1,Ibs2,Tot\}] [{\tt integer}] number of events to generate,
	\item [nCells\{Born,Virt,Hard,Brdq,Ibaa,Ibaq,Ibs1,Ibs2\}] [{\tt integer}] number of cells in buildup per branch,
	\item [nSampl\{Born,Virt,Hard,Brdq,Ibaa,Ibaq,Ibs1,Ibs2\}] [{\tt integer}] number of MC events per cell in build-up per branch,
	\item [nBin\{Born,Virt,Hard,Brdq,Ibaa,Ibaq,Ibs1,Ibs2\}] [{\tt integer}] number of bins for search for the optimal division per branch,
	\item [EvPerBin\{Born,Virt,Hard,Brdq,Ibaa,Ibaq,Ibs1,Ibs2\}] [{\tt integer}] maximum events (equiv.) per bin in buid-up per branch,
	\item [MaxWtRej\{Born,Virt,Hard,Brdq,Ibaa,Ibaq,Ibs1,Ibs2\}] [{\tt integer}] maximum weight for rejection per branch,
	\item [RanSeed] [{\tt integer}] pseudorandom generator seed.
\end{description}

Different contributions to the NLO cross sections (real, virtual) can differ from
each other by orders of magnitude. To improve relative and absolute
accuracy, a user can enlarge \textit{nCellsVirt}. In general, it also makes generation speed higher.

\subsubsection{Standard Model parameters}

\begin{description}
	\item [alpha, gf, alphas, conhc] [{\tt double}] a list of constants and coefficients: 
	\(\alpha_{EM}\), \(G_{\mu}\), \(\alpha_{S}\), conversion constant from GeV$^{-2}$ to pb,
	\item [mw, mz, mh] [{\tt double}] W, Z, Higgs boson masses,
	\item [wz, ww, wh, wtp] [{\tt double}] W, Z, Higgs and the top quark widths,
	\item [men, mel, mmn, mmo, mtn, mta] [{\tt double}] \(\nu_e,e,\nu_{\mu}, \mu, \nu_{\tau}, \tau\) lepton masses,
	\item [mdn, mup, mst, mch, mbt, mtp] [{\tt double}] \(d,u,s,c,b,t\) quark masses.
	
\end{description}

\section{Numerical results}
\label{sec:validation}
\subsection{Input paremeters}
For numerical validation we worked in the $\alpha(0)$ and $G_\mu$ schemes and used the following set of input parameters:
\begin{eqnarray}
&&\alpha^{-1}(0)= 137.035999084, \quad G_F = 1.1663787\times10^{-5} \; \text{GeV}^{-2}, \nonumber\\
&&M_W = 80.379 \; \text{GeV}, \quad M_Z = 91.1876 \; \text{GeV}, \nonumber\\
&&M_H = 125.25 \; \text{GeV}, \nonumber\\
&&\Gamma_W = 2.085 \; \text{GeV}, \quad \Gamma_Z = 2.4952 \; \text{GeV}, \nonumber\\
&&|V_{ud}| = 0.9737, \quad |V_{us}| = 0.2252, \nonumber\\
&&|V_{cd}| = 0.221, \quad |V_{cs}| = 0.987, \nonumber\\
&&|V_{cb}| = 0, \quad |V_{ub}| = 0, \nonumber\\
&&m_e = 0.51099895 \; \text{MeV}, \quad m_\mu = 0.1056583745 \; \text{GeV}, \nonumber\\
&&m_\tau = 1.77686 \; \text{GeV},\nonumber\\
&&m_d = 0.066 \; \text{GeV}, \quad m_u = 0.066 \; \text{GeV},\nonumber\\
&&m_s = 0.15 \; \text{GeV}, \quad m_c = 1.67 \; \text{GeV},\nonumber\\
&&m_b = 4.78 \; \text{GeV}, \quad m_t = 172.76 \; \text{GeV}.
\end{eqnarray}
The values of the parameters were taken from PDG-2020~\cite{ParticleDataGroup:2020ssz}, except for the masses of light quarks (u, d and s), for which the same values as in~\cite{Dittmaier:2009cr} were used. We used the {\tt CT14nlo} PDF set~\cite{Dulat:2015mca} for $qq$ and $qg$ subprocesses and {\tt CT14qed\_inc\_proton} PDF set~\cite{Schmidt:2015zda} for $q\gamma$ and $\gamma\gamma$ subprocesses via the LHAPDF6 library~\cite{Buckley:2014ana} with the factorization and the renormalization scales $M_{\ell\ell}$ for NC DY and $M_{\ell\nu}$ for CC DY.

\subsection{Cuts}

The following cuts were applied ($\ell = e, \; \mu$):
\begin{eqnarray}
\label{Cuts}
&&pp[p\bar{p}] \to \ell^+\ell^-X: \quad p_\perp(\ell^\pm) > 25 \; \text{GeV}, \quad |\eta(\ell^\pm)| < 2.5, \quad M(\ell^+\ell^-) > 50 \; \text{GeV}, \nonumber \\
&&pp[p\bar{p}] \to \ell^+\nu_\ell X: \quad p_\perp(\ell^+) > 25 \; \text{GeV}, \quad p_\perp(\nu_\ell) > 25 \; \text{GeV}, \quad |\eta(\ell^+)| < 2.5, \quad M(\ell^+\nu_\ell) > 1 \; \text{GeV}, \nonumber \\
&&pp[p\bar{p}] \to \ell^-\bar{\nu}_\ell X: \quad p_\perp(\ell^-) > 25 \; \text{GeV}, \quad p_\perp(\bar{\nu}_\ell) > 25 \; \text{GeV}, \quad |\eta(\ell^-)| < 2.5, \quad M(\ell^-\bar{\nu}_\ell) > 1 \; \text{GeV}.
\end{eqnarray}

\subsection{List of observables}

\subsubsection{Z boson observables}

\begin{figure}[!h]
    \centering
    \includegraphics{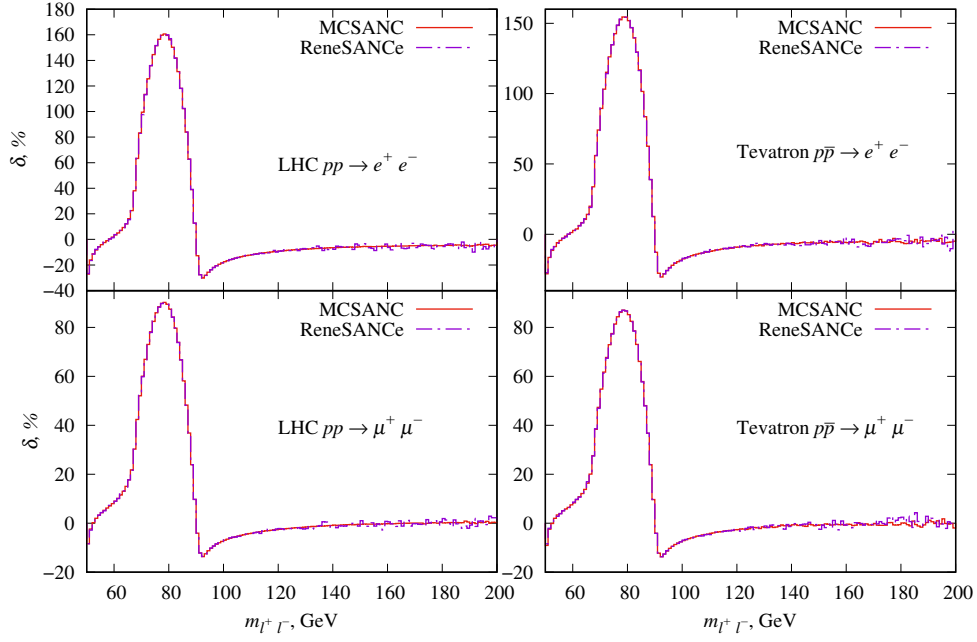}
    \caption{The lepton pair invariant mass $m_{l^+l^-}$ distributions for relative corrections.}
    \label{fig:NC-gf0-minv}
\end{figure}

\begin{figure}[!h]
    \centering
    \includegraphics{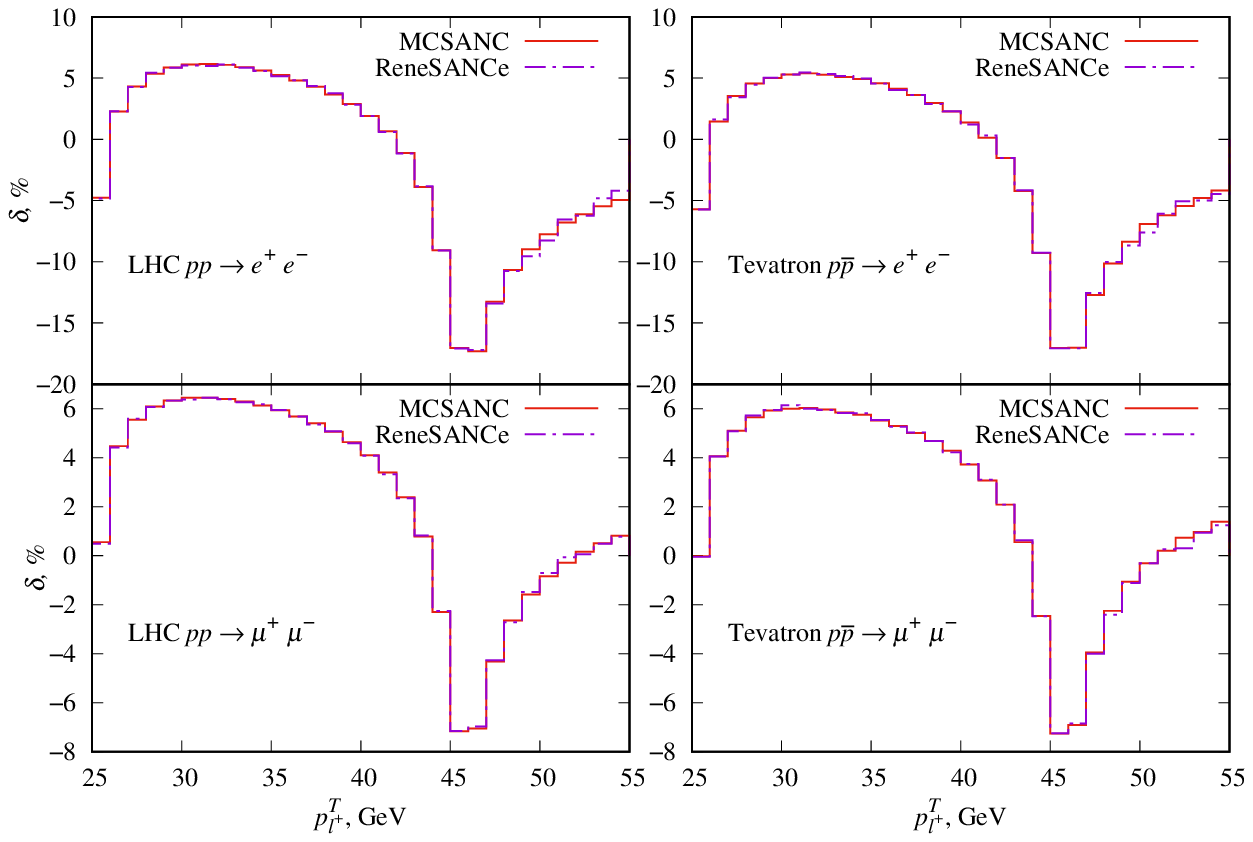}
    \caption{The lepton transverse momentum $p^T_{l^+}$ distributions for relative corrections.}
    \label{fig:NC-gf0-pt3}
\end{figure}

\begin{figure}[!h]
    \centering
    \includegraphics{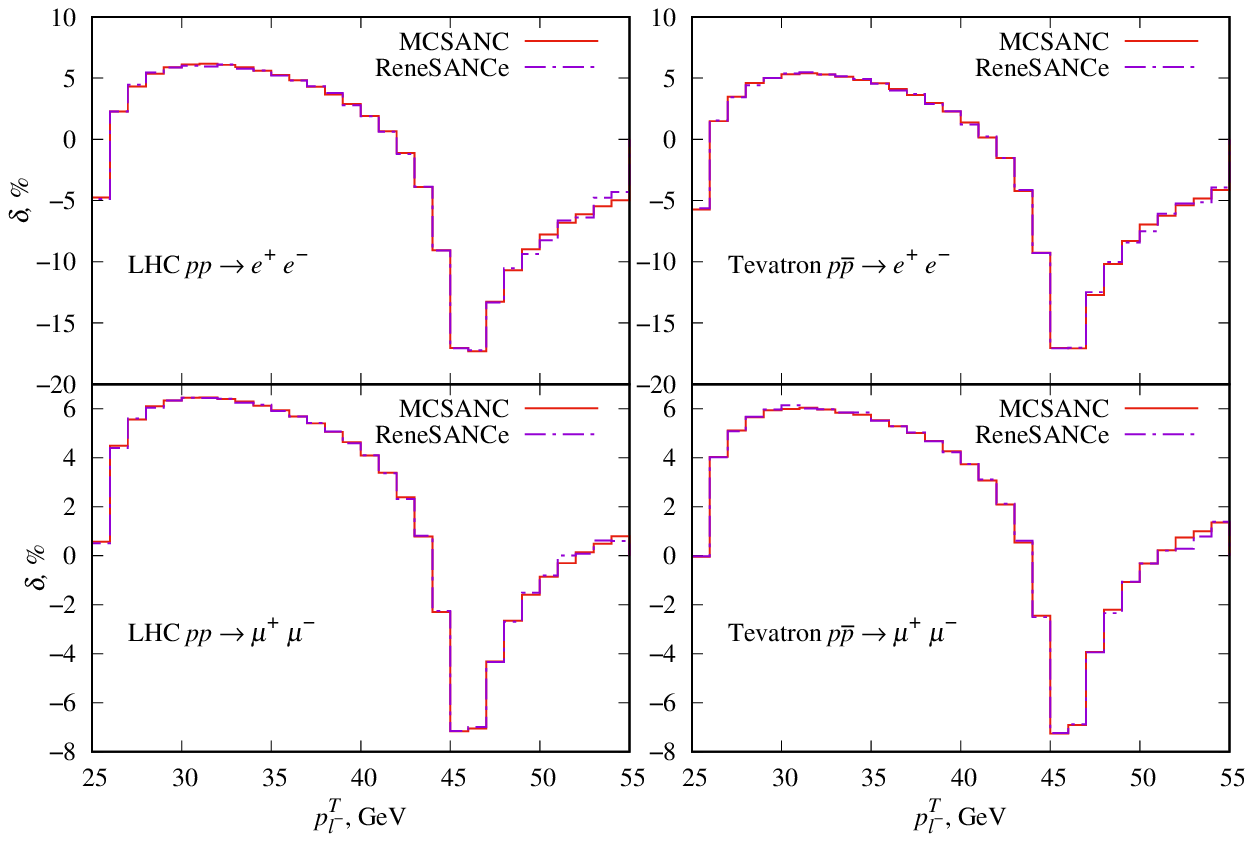}
    \caption{The antilepton transverse momentum $p^T_{l^-}$ distributions for relative corrections.}
    \label{fig:NC-gf0-pt4}
\end{figure}

\begin{figure}[!h]
    \centering
    \includegraphics{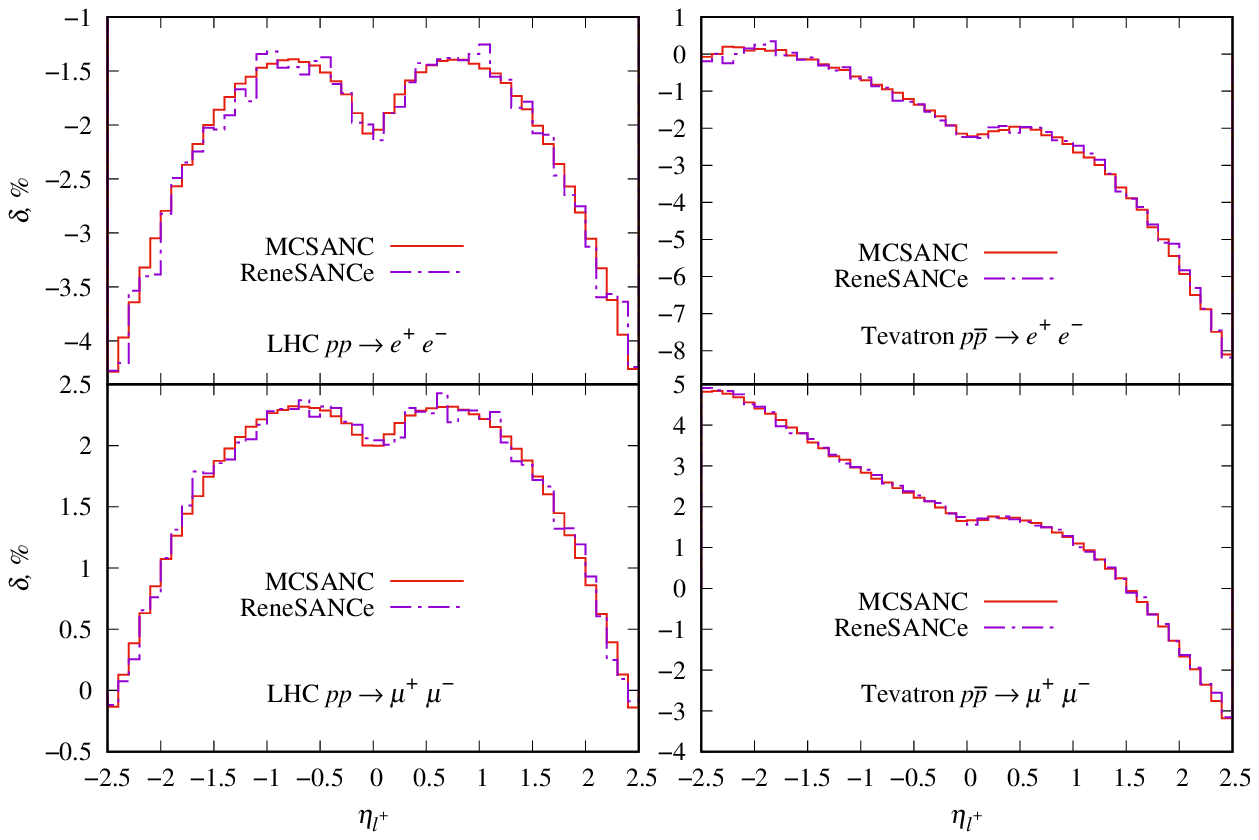}
    \caption{The lepton pseudorapidity $\eta_{l^+}$ distributions for relative corrections.}
    \label{fig:NC-gf0-eta3}
\end{figure}

\begin{figure}[!h]
    \centering
    \includegraphics{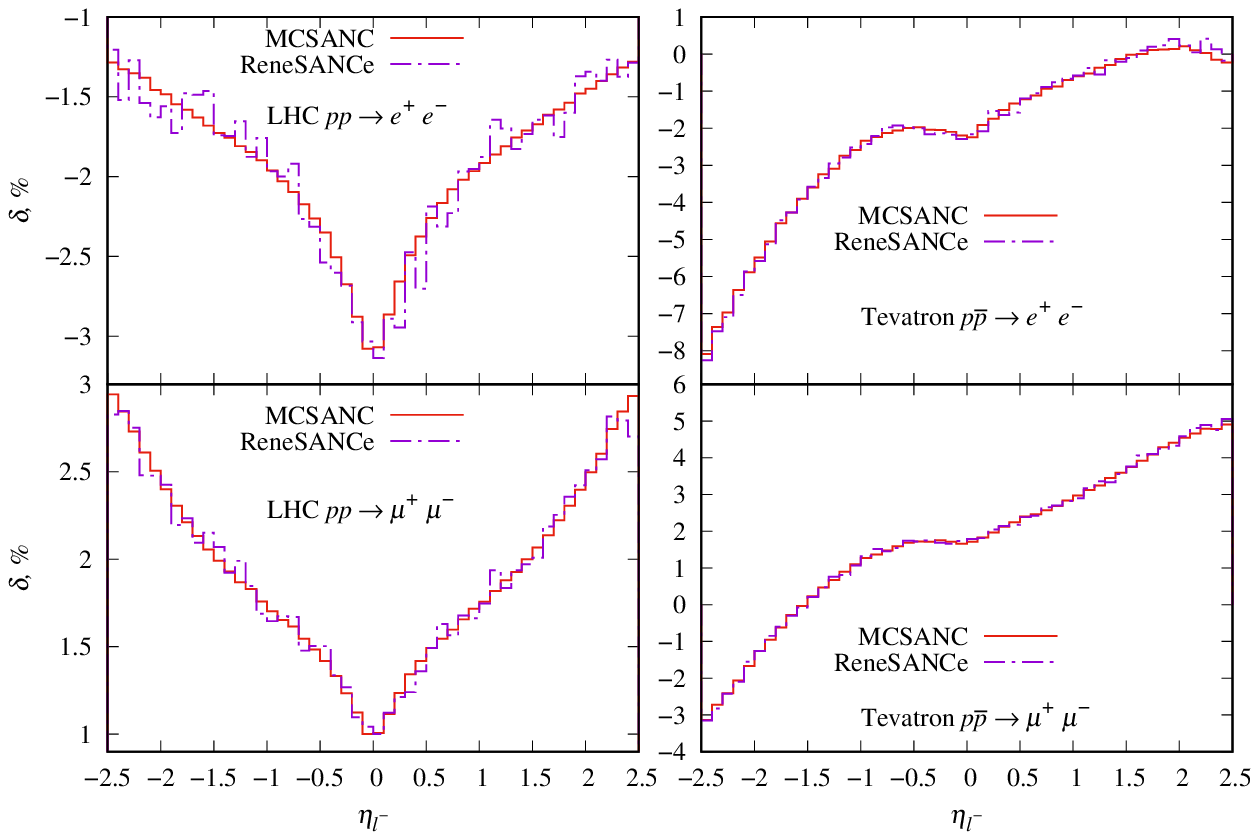}
    \caption{The lepton pseudorapidity $\eta_{l^-}$ distributions for relative corrections.}
    \label{fig:NC-gf0-eta4}
\end{figure}

\begin{figure}[!h]
    \centering
    \includegraphics{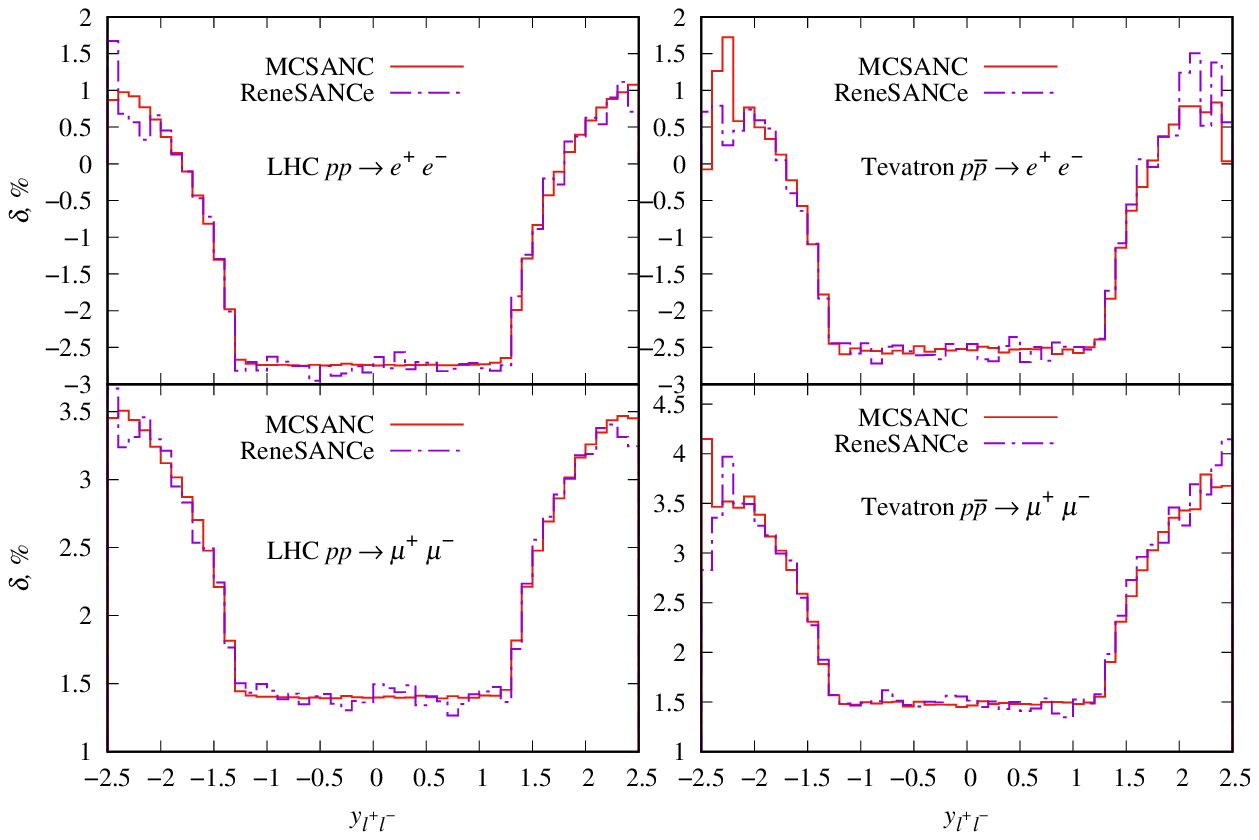}
    \caption{The lepton pair rapidity $y_{l^+l^-}$ distributions for relative corrections.}
    \label{fig:NC-gf0-y34}
\end{figure}

\begin{figure}[!h]
    \centering
    \includegraphics{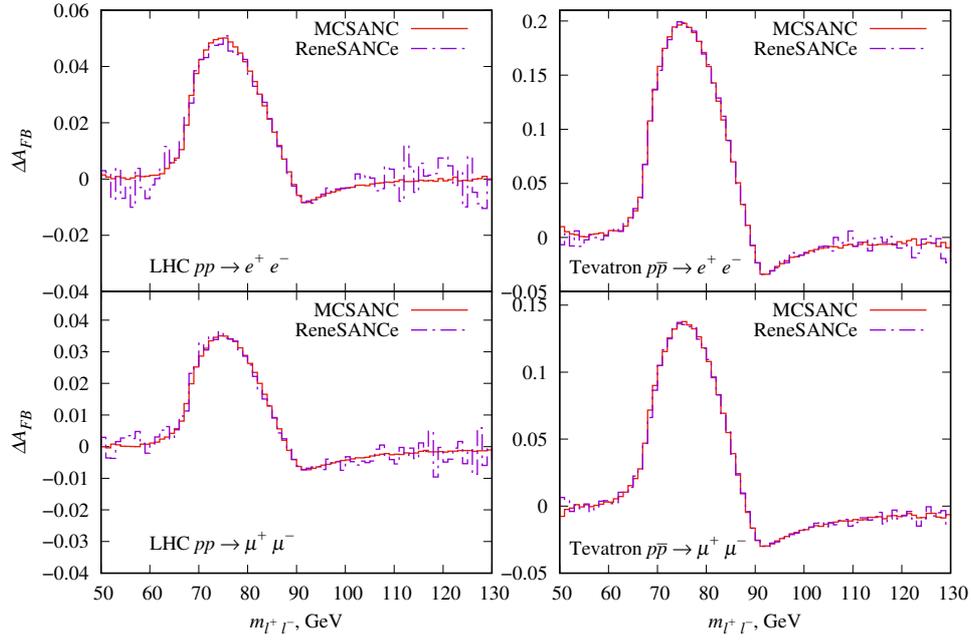}
    \caption{The forward-backward asymmetry $A_{\rm FB}$ distributions.}
    \label{fig:NC-gf0-dafb}
\end{figure}

\begin{itemize}
\item{$\sigma_{\sss Z}$:~total inclusive cross section of $Z$ boson production}.

The total inclusive  cross section $\sigma=\sigma_{\sss Z}$ and the corresponding relative corrections $\delta$ for the LHC in the $\alpha(0)$ scheme are shown in Tables
(\ref{tab:LHC8TeVgf0ew}, \ref{tab:LHC8TeVgf0phot}), and in the $G_\mu$ scheme in Tables (\ref{tab:LHC8TeVgf1ew}, \ref{tab:LHC8TeVgf1phot}),
and the same results for the Tevatron in the $\alpha(0)$ in
Tables (\ref{tab:Tevatron1_96TeVgf0}, \ref{tab:Tevatron1_96TeVPhotgf0})
and $G_\mu$ scheme in Tables (\ref{tab:Tevatron1_96TeVgf1}, \ref{tab:Tevatron1_96TeVPhotgf1}).

\item{$\frac{\ds d \sigma}{\ds d m_{\ell^+\ell^-}}$: invariant mass distribution of the final-state lepton pair.}

The relative corrections $\delta$ for $m_{\ell^+\ell^-}$ range $[50 \text{ GeV}; 200 \text{ GeV}]$ are shown in Fig.~\ref{fig:NC-gf0-minv}
for LHC at $\sqrt{s_0} = 8$ TeV (left)
and for Tevatron at 
$\sqrt{s_0} = 1.96$ TeV (right).

\item{
$\frac{\ds d\sigma}{\ds dp^T_{\ell^+}}$:~
transverse momentum distribution of the $\ell^+$.}

The relative corrections $\delta$ for $p^T_{\ell^+}$ range $[25 \text{ GeV}; 55 \text{ GeV}]$
are shown in Fig.\ref{fig:NC-gf0-pt3}.

\item{
$\frac{\ds d\sigma}{\ds dp^T_{\ell^-}}$:~
transverse momentum distribution of the $\ell^-$.}

The relative corrections $\delta$ for $p^T_{\ell^-}$ range $[25 \text{ GeV}; 55 \text{ GeV}]$
are shown in Fig.\ref{fig:NC-gf0-pt4}.

\item{
$\frac{\ds d\sigma}{\ds d\eta_{\ell^+}}$:
~pseudorapidity distribution of the $\ell^+$.}

The relative corrections $\delta$ for $\eta_{\ell^+}$ range $[-2.5; 2.5]$ are shown in Fig.\ref{fig:NC-gf0-eta3}.

\item{
$\frac{\ds d \sigma}{\ds d\eta_{\ell^-}}$:
~pseudorapidity distribution of the $\ell^-$.}

The relative corrections $\delta$ for $\eta_{\ell^-}$ range $[-2.5; 2.5]$ are shown in Fig.\ref{fig:NC-gf0-eta4}.

\item{
$\frac{\ds d \sigma}{\ds dy_{\ell^+\ell^-}}$:~rapidity distribution of the final-state lepton pair.}

The relative corrections $\delta$ for $y_{\ell^+\ell^-}$ range $[-2.5; 2.5]$ are shown in Fig.\ref{fig:NC-gf0-y34}.

\item{$A_{\rm FB}$:~forward-backward asymmetry as a function of $m_{\ell^+\ell^-}$.}

The difference $\Delta A_{FB} = A_{FB}^{NLO} - A_{FB}^{LO}$ for $m_{\ell^+\ell^-}$ range $[50 \text{ GeV}; 130 \text{ GeV}]$  are shown in Fig.\ref{fig:NC-gf0-dafb}.

$A_{FB}$ is defined by
\begin{eqnarray}
A_{FB}=\frac{\sigma_F-\sigma_B}{\sigma_F+\sigma_B},
\end{eqnarray}
where $\sigma_F=\int^1_0 \frac{\ds d\sigma}{\ds d\cos\vartheta^*} d\cos\vartheta^*$,
and $\sigma_B=\int^0_{-1} \frac{\ds d\sigma}{\ds d\cos\vartheta^*} d\cos\vartheta^*$.
The definition for $\cos\vartheta^*$ is as follows:
\begin{equation}
\cos \vartheta^*=F
\frac{\ds 2}{\ds m_{\ell^+ \ell^-}\sqrt{m^2_{\ell^+\ell^-}+\left(p^T_{\ell^+\ell^-}\right)^2}}
[p^+_{\ell^-}p^-_{\ell^+}-p^-_{\ell^-}p^+_{\ell^+}]
\end{equation}
with
\begin{equation}
    p^\pm = \frac{\ds 1}{\ds\sqrt{2}}(E\pm p^z).
\end{equation}
Here $F = 1$ for $p\bar{p}$ collisions, and $F = \frac{\ds |p^z_{\ell^+\ell^-}|}{\ds p^z_{\ell^+\ell^-}}$ for $pp$ collisions.
\end{itemize}

\subsubsection{W boson observables}

\begin{figure}[!h]
    \centering
    \includegraphics{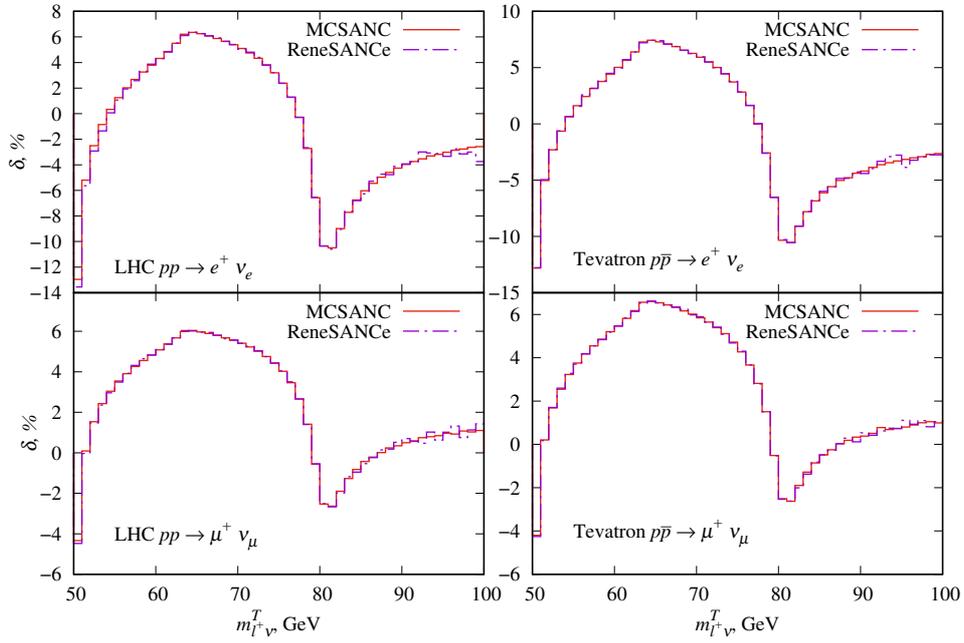}
    \caption{The lepton pair transverse mass $m^T_{l^+\nu}$ distributions for relative corrections.}
    \label{fig:CC-Wp-gf0-mtr}
\end{figure}

\begin{figure}[!h]
    \centering
    \includegraphics{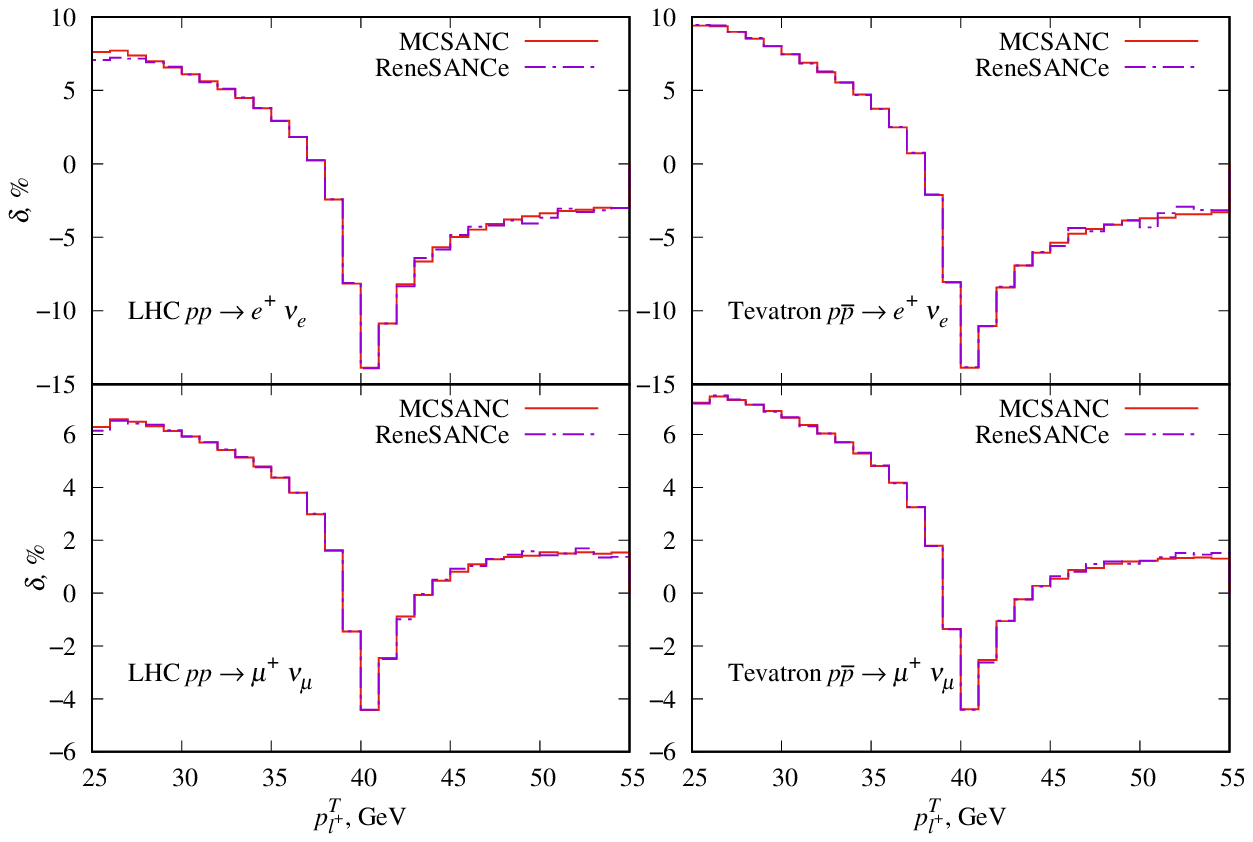}
    \caption{The lepton transverse momentum $p^{T}_{l^+}$ distributions for relative corrections.}
    \label{fig:CC-Wp-gf0-pt3}
\end{figure}

\begin{figure}[!h]
    \centering
    \includegraphics{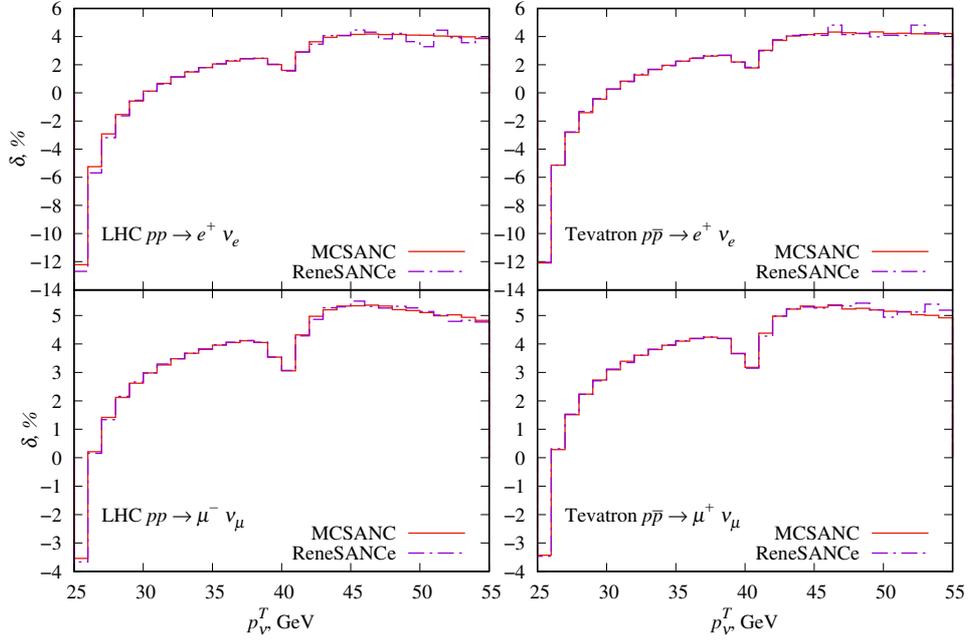}
    \caption{The neutrino transverse momentum $p^T_\nu$ distributions for relative corrections.}
    \label{fig:CC-Wp-gf0-pt4}
\end{figure}

\begin{figure}[!h]
    \centering
    \includegraphics{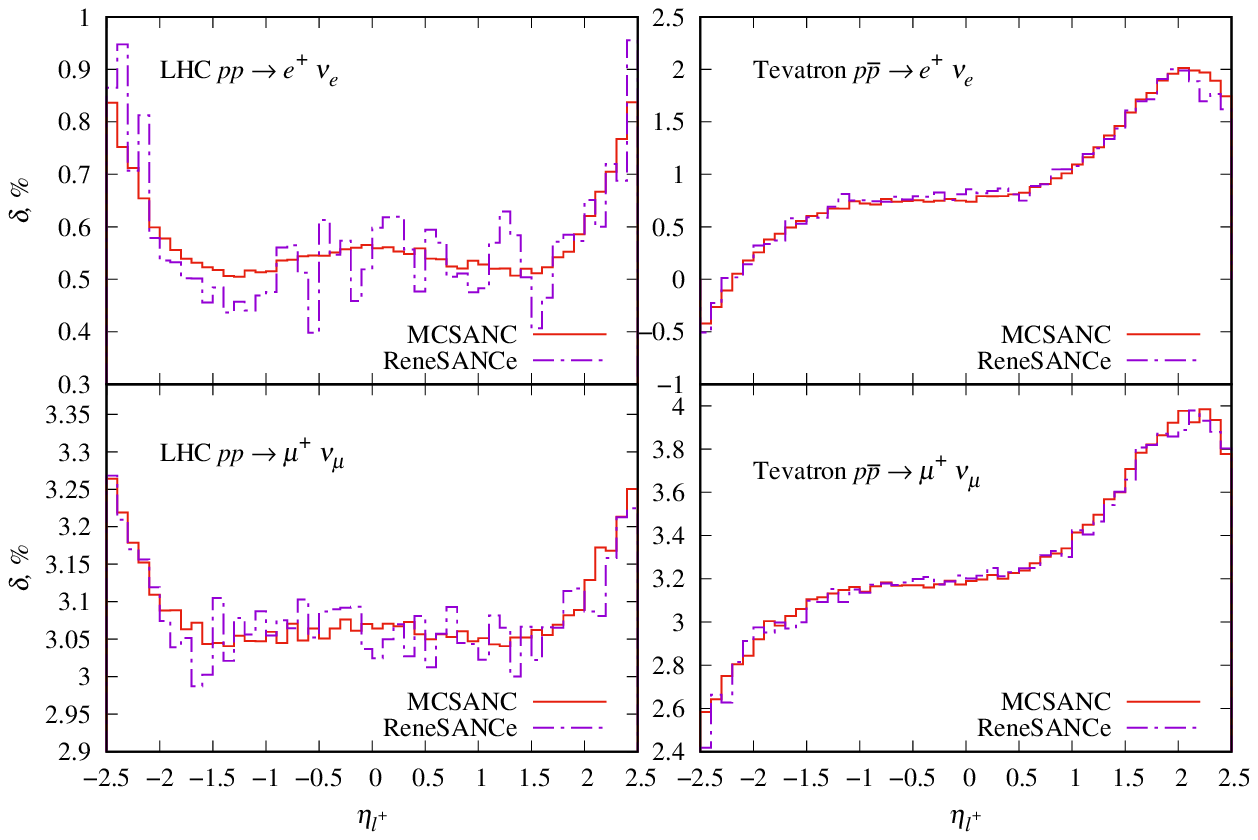}
    \caption{The antilepton pseudorapidity $\eta_{l^+}$ distributions for relative corrections.}
    \label{fig:CC-Wp-gf0-eta3}
\end{figure}

\begin{figure}[!h]
    \centering
    \includegraphics{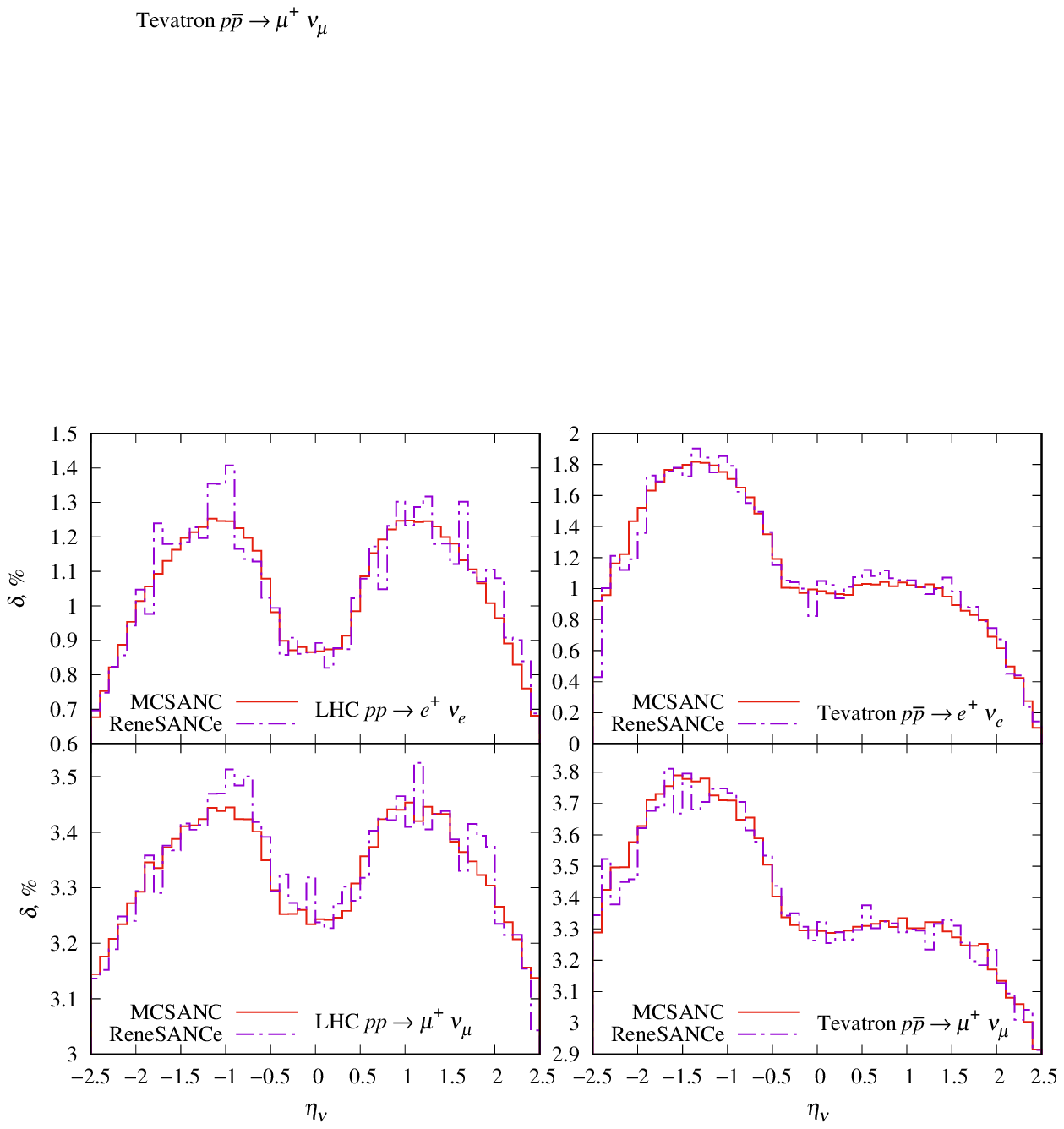}
    \caption{The neutrino pseudorapidity $\eta_\nu$ distributions for relative corrections.}
    \label{fig:CC-Wp-gf0-eta4}
\end{figure}

\begin{figure}[!h]
    \centering
    \includegraphics{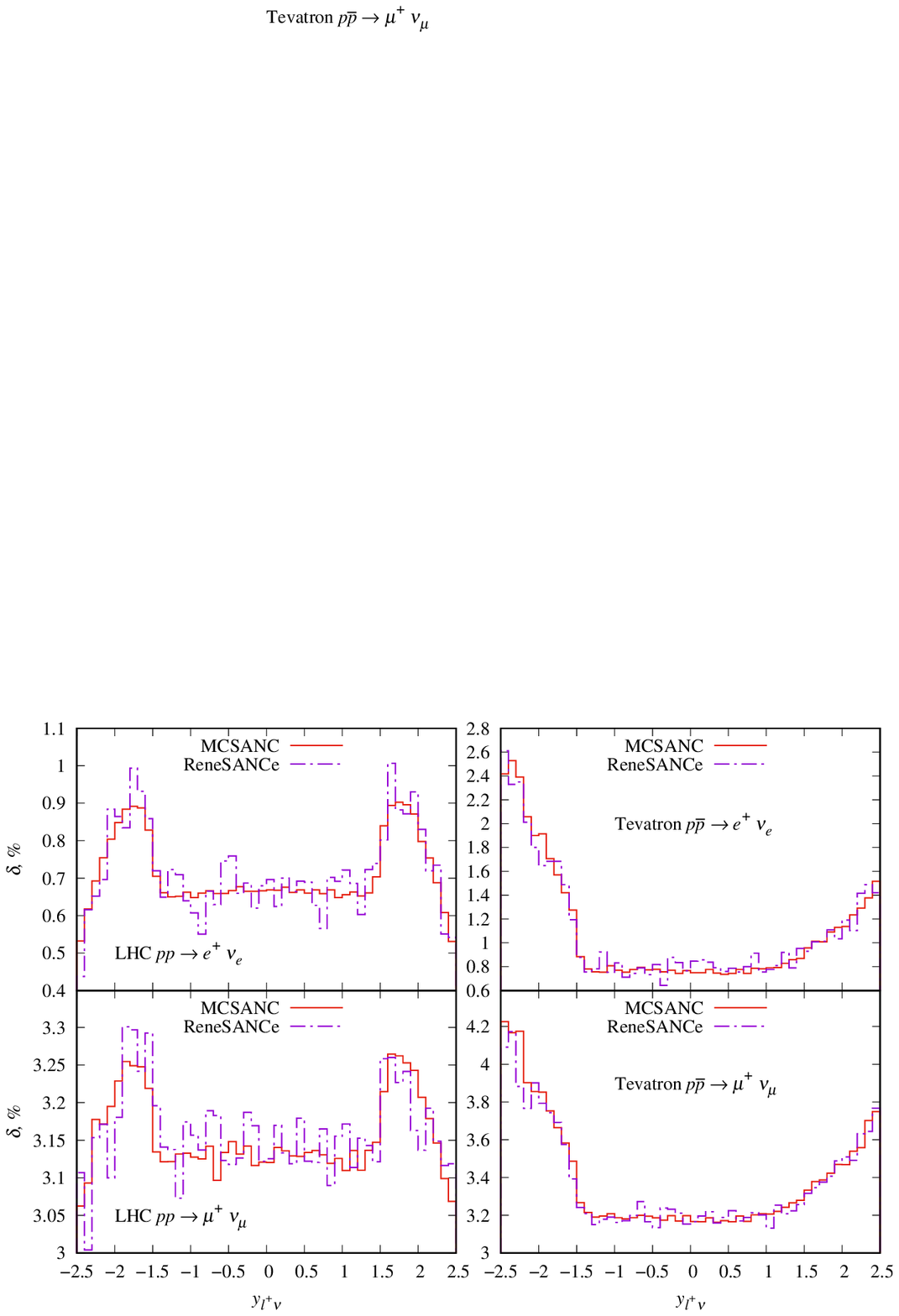}
    \caption{The lepton pair rapidity $y_{l^+\nu}$ distributions for relative corrections.}
    \label{fig:CC-Wp-gf0-y34}
\end{figure}

\begin{figure}[!h]
    \centering
    \includegraphics{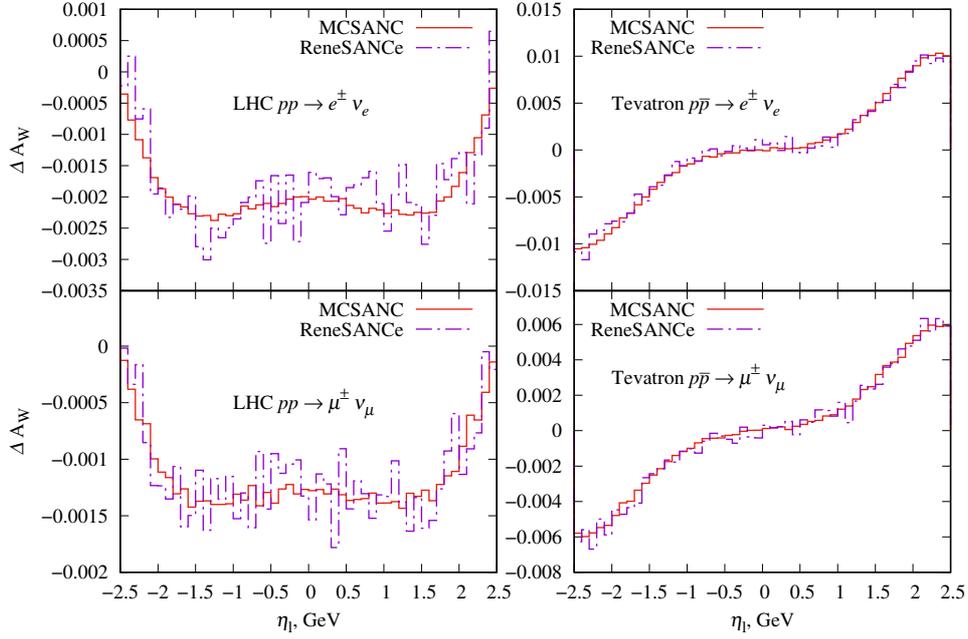}
    \caption{The charge asymmetry on antilepton pseudorapidity $\eta_{l^+}$ distributions for relative NLO and LO difference.}
    \label{fig:CC-asymm-gf0-eta3}
\end{figure}

\begin{figure}[!h]
    \centering
    \includegraphics{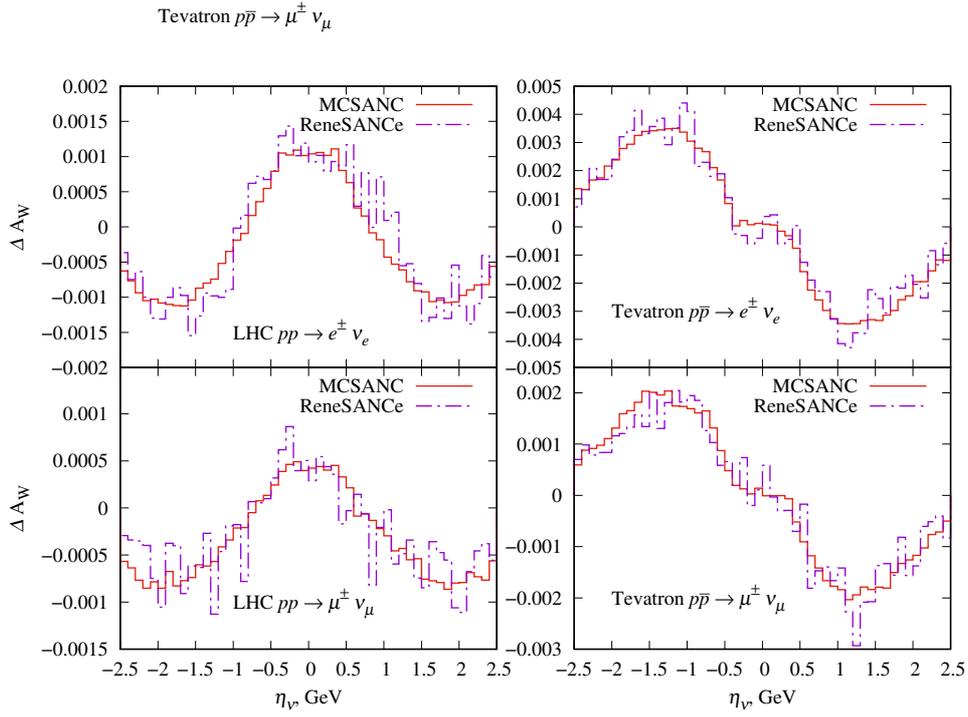}
    \caption{The charge asymmetry on neutrino pseudorapidity $\eta_\nu$ distributions for relative NLO and LO difference.}
    \label{fig:CC-asymm-gf0-eta4}
\end{figure}

\begin{figure}[!h]
    \centering
    \includegraphics{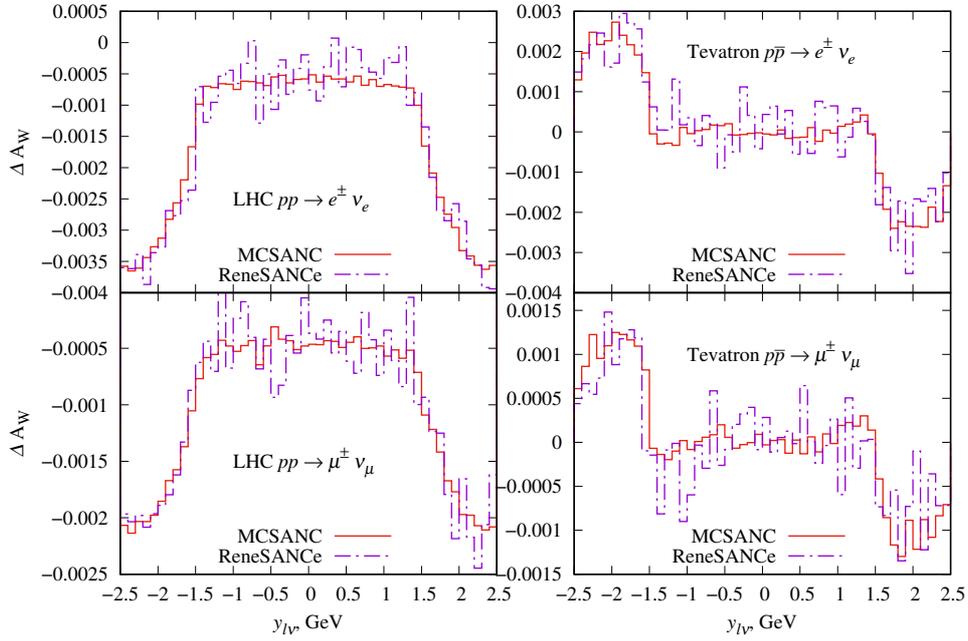}
    \caption{The charge asymmetry on lepton-neutrino pair rapidity $y_{l\nu}$ distributions for relative NLO and LO difference.}
    \label{fig:CC-asymm-gf0-y34}
\end{figure}

\begin{itemize}
\item{$\sigma_{\sss W}$:~total inclusive cross section of $W$ boson production}

\item{$\frac{\ds d \sigma}{\ds d m^T_{\ell^+\nu_{\ell}}}$: 
transverse mass distribution of the final state lepton-neutrino pair.}

\begin{equation}
m^T_{\ell^+\nu_{\ell}}=\sqrt{2p^T_{\ell^+}p^T_{\nu_\ell} (1-\cos \phi_{\ell^+\nu_\ell})},
\end{equation}
where $p^T_{\ell^+}$, $p^T_{\nu_\ell}$ are the transverse momenta of charged lepton and the neutrino, and $\phi_{\ell^+\nu_\ell}$ is the angle between the charged lepton and the neutrino in the transverse plane.

The relative corrections $\delta$ for $m^T_{\ell^+\nu_{\ell}}$ range $[50 \text{ GeV}; 100 \text{ GeV}]$ are shown in Fig.\ref{fig:CC-Wp-gf0-mtr}.

\item{
$\frac{\ds d\sigma}{\ds dp^T_{\ell^+}}$:~
transverse momentum distribution of the $\ell^+$.}

The relative corrections $\delta$ for $p^T_{\ell^+}$ range $[25 \text{ GeV}; 55 \text{ GeV}]$
are shown in Fig.\ref{fig:CC-Wp-gf0-pt3}.

\item{
$\frac{\ds d\sigma}{\ds dp^T_{\nu_\ell}}$:~
transverse momentum distribution of the neutrino $\nu_\ell$.}

The relative corrections $\delta$ for $p^T_{\nu_\ell}$ range $[25 \text{ GeV}; 55 \text{ GeV}]$
are shown in Fig.\ref{fig:CC-Wp-gf0-pt4}.

\item{
$\frac{\ds d\sigma}{\ds d\eta_{\ell^+}}$:
~pseudorapidity distribution of the $\ell^+$.}

The relative corrections $\delta$ for $\eta_{\ell^+}$ range $[-2.5; 2.5]$ are shown in Fig.\ref{fig:CC-Wp-gf0-eta3}.

\item{
$\frac{\ds d \sigma}{\ds d\eta_{\nu_\ell}}$:
~pseudorapidity distribution of the neutrino $\nu_\ell$.}

The relative corrections $\delta$ for $\eta_{\nu_\ell}$ range $[-2.5; 2.5]$ are shown in Fig.\ref{fig:CC-Wp-gf0-eta4}.

\item{
$\frac{\ds d \sigma}{\ds dy_{\ell^+\ell^-}}$:~rapidity distribution of the final-state lepton-neutrino pair.}

The relative corrections $\delta$ for $y_{\ell^+\nu_\ell}$ range $[-2.5; 2.5]$ are shown in Fig.\ref{fig:CC-Wp-gf0-y34}.

\item{$A_W$: the charge asymmetry.}

$A_W$ is defined by
\begin{eqnarray}
A_{W}=\frac{\sigma^{W^+}-\sigma^{W^-}}{\sigma^{W^+}+\sigma^{W^-}}.
\end{eqnarray}

The difference $\Delta A_{W} = A_{W}^{NLO} - A_{W}^{LO}$ as a function of charged lepton pseudorapidity $\eta_\ell$ in range $[-2.5; 2.5]$ is shown in Fig.\ref{fig:CC-asymm-gf0-eta3}.

$\Delta A_{W}$ as a function of neutrino pseudorapidity $\eta_{\nu_\ell}$ in range $[-2.5; 2.5]$ is shown in Fig.\ref{fig:CC-asymm-gf0-eta4}.

$\Delta A_{W}$ as a function of a rapidity $y_{\ell\nu_\ell}$ of the lepton-neutrino pair in range $[-2.5; 2.5]$ is shown in Fig.\ref{fig:CC-asymm-gf0-y34}.
\end{itemize}

\begin{table}[ht!]
    \centering
    \begin{tabular}{|r|l|l|l|l|l|l|}
        \hline
        $pp \to $ & $e^+e^- X$ & $e^+\nu_e X$ & $e^-\bar{\nu}_e X$ & $\mu^+\mu^- X$ & $\mu^+\nu_\mu X$ & $\mu^-\bar{\nu}_\mu X$ \\
        \hline
        \multirow{2}{*}{$\sigma_{\text{LO}}$ [pb]} \; 
        {\tt ReneSANCe} & 427.78(1) & 2903.7(1) & 1979.5(1) & 427.78(1) & 2903.7(1) & 1979.4(1)\\
        {\tt MCSANC}    & 427.78(1) & 2903.7(1) & 1979.5(1) & 427.78(1) & 2903.7(1) & 1979.5(1)\\
        \hline
        \multirow{2}{*}{$\Delta\sigma_{\text{NLO EW}}$ [pb]} \; 
        {\tt ReneSANCe} & $-8.64(1)$ & 16.6(1) & 19.4(1) & 7.70(1) & 89.6(1) & 66.1(1)\\
        {\tt MCSANC}    & $-8.62(1)$ & 16.6(1) & 19.3(1) & 7.71(1) & 89.5(2) & 66.0(1)\\
        \hline
        \multirow{2}{*}{$\Delta\sigma_{\text{NLO QCD}}$ [pb]} \; 
        {\tt ReneSANCe} & 26.1(1) & $-5.8(6)$ & 36.2(4) & & & \\
        {\tt MCSANC}    & 26.1(1) & $-5.7(3)$  & 36.8(3) & & & \\
        \hline
        \multirow{2}{*}{$\delta_{\text{NLO EW}}$ [\%]} \; 
        {\tt ReneSANCe} & $-2.02(1)$ & 0.57(1) & 0.98(1) & 1.80(1) & 3.09(1) & 3.34(1)\\
        {\tt MCSANC}    & $-2.02(1)$ & 0.57(1) & 0.98(1) & 1.80(1) & 3.08(1) & 3.33(1)\\
        \hline
        \multirow{2}{*}{$\delta_{\text{NLO QCD}}$ [\%]} \; 
        {\tt ReneSANCe} & 6.11(1) & $-0.20(2)$ & 1.83(2) & 7.25(3) & 0.14(3) & 2.64(3)\\
        {\tt MCSANC}    & 6.11(1) & $-0.19(1)$ & 1.86(1) &  & \\
        \hline
    \end{tabular}
    \caption{Cross sections and relative corrections for NC and CC DY processes at $\sqrt{s_0} = 8$ TeV in $\alpha(0)$-scheme obtained by {\tt ReneSANCe v1.3.0} and {\tt MCSANC v1.21}.}
    \label{tab:LHC8TeVgf0ew}
\end{table}

\begin{table}[ht!]
    \centering
    \begin{tabular}{|r|l|l|l|l|l|l|}
        \hline
        $pp \to $ & $e^+e^- X$ & $e^+\nu_e X$ & $e^-\bar{\nu}_e X$ & $\mu^+\mu^- X$ & $\mu^+\nu_\mu X$ & $\mu^-\bar{\nu}_\mu X$ \\
        \hline
        \multirow{2}{*}{$\sigma_{\text{LO}}$ [pb]} \; 
        {\tt ReneSANCe} & 423.53(1) & 2877.1(1) & 1957.7(1) & 423.52(1) & 2877.2(1) & 1957.7(1)\\
        {\tt MCSANC}    & 423.53(1) & 2877.2(1) & 1957.7(1) & 423.53(1) & 2877.2(1) & 1957.7(1) \\
        \hline
        \multirow{2}{*}{$\Delta\sigma_{\gamma q}$ [pb]} \; 
        {\tt ReneSANCe} & $-$0.364(1) & 1.098(1) & 0.799(1) & $-$0.372(1) & 1.098(1) & 0.799(1)\\
        {\tt MCSANC}    & $-$0.364(1) & 1.098(1) & 0.799(1) & $-$0.371(1) & 1.099(1) & 0.800(1) \\
        \hline
        \multirow{2}{*}{$\Delta\sigma_{\gamma \gamma}$ [pb]} \; 
        {\tt ReneSANCe} & 0.648(1) & $-$ & $-$ & 0.648(1) & $-$ & $-$\\
        {\tt MCSANC}    & 0.649(1) & $-$ & $-$ & 0.648(1) & $-$ & $-$\\
        \hline
        \multirow{2}{*}{$\delta_{\gamma q}$ [\%]} \; 
        {\tt ReneSANCe} & $-$0.086(1) & 0.038(1) & 0.041(1) & $-$0.088(1) & 0.038(1) & 0.041(1)\\
        {\tt MCSANC}    & $-$0.086(1) & 0.038(1) & 0.041(1) & $-$0.088(1) & 0.038(1) & 0.041(1)\\
        \hline
        \multirow{2}{*}{$\delta_{\gamma \gamma}$ [\%]} \; 
        {\tt ReneSANCe} & 0.153(1) & $-$ & $-$ & 0.153(1) & $-$ & $-$\\
        {\tt MCSANC}    & 0.153(1) & $-$ & $-$ & 0.153(1) & $-$ & $-$\\
        \hline
    \end{tabular}
    \caption{Cross sections and relative corrections for NC and CC DY processes at $\sqrt{s_0} = 8$ TeV in $\alpha(0)$-scheme obtained by {\tt ReneSANCe v1.3.0} and {\tt MCSANC v1.21}.}
    \label{tab:LHC8TeVgf0phot}
\end{table}

\begin{table}[ht!]
    \centering
    \begin{tabular}{|r|l|l|l|l|l|l|}
        \hline
        $pp \to $ & $e^+e^- X$ & $e^+\nu_e X$ & $e^-\bar{\nu}_e X$ & $\mu^+\mu^- X$ & $\mu^+\nu_\mu X$ & $\mu^-\bar{\nu}_\mu X$ \\
        \hline
        \multirow{2}{*}{$\sigma_{\text{LO}}$ [pb]} \; 
        {\tt ReneSANCe} & 459.75(1) & 3120.8(1) & 2127.5(1) & 459.76(1) & 3120.8(1) & 2127.4(1)\\
        {\tt MCSANC}    & 459.77(1) & 3120.8(1) & 2127.3(1) & 459.76(1) & 3120.8(1) & 2127.3(1)\\
        \hline
        \multirow{2}{*}{$\Delta\sigma_{\text{NLO EW}}$ [pb]} \; 
        {\tt ReneSANCe} & $-36.29(2)$ & $-165.6(1)$ & $-104.4(1)$ & $-18.77(1)$ & $-87.3(1)$ & $-54.1(1)$\\
        {\tt MCSANC}    & $-36.30(1)$ & $-165.8(1)$ & $-104.4(1)$ & $-18.73(1)$ & $-87.3(1)$ & $-54.1(2)$\\
        \hline
        \multirow{2}{*}{$\delta_{\text{NLO EW}}$ [\%]} \; 
        {\tt ReneSANCe} & $-7.89(1)$ & $-5.31(1)$ & $-4.91(1)$ & $-4.08(1)$ & $-2.80(1)$ & $-2.54(1)$\\
        {\tt MCSANC}    & $-7.90(1)$ & $-5.31(1)$ & $-4.91(1)$ & $-4.07(1)$ & $-2.80(1)$ & $-2.54(1)$\\
        \hline
    \end{tabular}
    \caption{Cross sections and relative corrections for NC and CC DY processes at $\sqrt{s_0} = 8$ TeV in $G_\mu$-scheme obtained by {\tt ReneSANCe v1.3.0} and {\tt MCSANC v1.21}.}
    \label{tab:LHC8TeVgf1ew}
\end{table}

\begin{table}[ht!]
    \centering
    \begin{tabular}{|r|l|l|l|l|l|l|}
        \hline
        $pp \to $ & $e^+e^- X$ & $e^+\nu_e X$ & $e^-\bar{\nu}_e X$ & $\mu^+\mu^- X$ & $\mu^+\nu_\mu X$ & $\mu^-\bar{\nu}_\mu X$ \\
        \hline
        \multirow{2}{*}{$\sigma_{\text{LO}}$ [pb]} \; 
        {\tt ReneSANCe} & 455.19(1) & 3092.2(1) & 2104.0(1) & 455.18(1) & 3092.2(1) & 2104.0(1)\\
        {\tt MCSANC}    & 455.19(1) & 3092.3(1) & 2104.0(1) & 455.19(1) & 3092.3(1) & 2104.0(1)\\
        \hline
        \multirow{2}{*}{$\Delta\sigma_{\gamma q}$ [pb]} \; 
        {\tt ReneSANCe} & $-0.391(1)$ & 1.180(1) & 0.858(1) & $-0.399(1)$ & 1.180(1) & 0.859(1) \\
        {\tt MCSANC}    & $-0.392(1)$ & 1.180(1) & 0.859(1) & $-0.399(1)$ & 1.180(1) & 0.859(1) \\
        \hline
        \multirow{2}{*}{$\Delta\sigma_{\gamma \gamma}$ [pb]} \; 
        {\tt ReneSANCe} & 0.648(1) & $-$ & $-$ & 0.648(1) & $-$ & $-$\\
        {\tt MCSANC}    & 0.648(1) & $-$ & $-$ & 0.648(1) & $-$ & $-$\\
        \hline
        \multirow{2}{*}{$\delta_{\gamma q}$ [\%]} \; 
        {\tt ReneSANCe} & $-0.086(1)$ & 0.038(1) & 0.041(1) & $-0.088(1)$ & 0.038(1) & 0.041(1)\\
        {\tt MCSANC}    & $-0.086(1)$ & 0.038(1) & 0.041(1) & $-0.088(1)$ & 0.038(1) & 0.040(1)\\
        \hline
        \multirow{2}{*}{$\delta_{\gamma \gamma}$ [\%]} \; 
        {\tt ReneSANCe} & 0.142(1) & $-$ & $-$ & 0.142(1) & $-$ & $-$\\
        {\tt MCSANC}    & 0.142(1) & $-$ & $-$ & 0.141(1) & $-$ & $-$\\
        \hline
    \end{tabular}
    \caption{Cross sections and relative corrections for NC and CC DY processes at $\sqrt{s_0} = 8$ TeV in $G_\mu$-scheme obtained by {\tt ReneSANCe v1.3.0} and {\tt MCSANC v1.21}.}
    \label{tab:LHC8TeVgf1phot}
\end{table}

\begin{table}[ht!]
    \centering
    \begin{tabular}{|r|l|l|l|l|l|l|}
        \hline
        $p\bar{p} \to $ & $e^+e^- X$ & $e^+\nu_e X$ & $e^-\bar{\nu}_e X$ & $\mu^+\mu^- X$ & $\mu^+\nu_\mu X$ & $\mu^-\bar{\nu}_\mu X$ \\
        \hline
        \multirow{2}{*}{$\sigma_{\text{LO}}$ [pb]} \; 
        {\tt ReneSANCe} & 134.17(1) & 683.15(1) & 683.15(1) & 134.16(1) & 683.14(1) & 683.14(2)\\
        {\tt MCSANC}    & 134.16(1) & 683.15(1) & 683.16(1) & 134.15(1) & 683.15(1) & 683.15(1)\\
        \hline
        \multirow{2}{*}{$\Delta\sigma_{\text{NLO EW}}$ [pb]} \; 
        {\tt ReneSANCe} & -2.75(1) & 6.15(1) & 6.15(1) & 2.38(1) & 22.39(1) & 22.40(1)\\
        {\tt MCSANC}    & -2.75(1) & 6.14(1) & 6.14(1) & 2.38(1) & 22.40(1) & 22.39(1)\\
        \hline
        \multirow{2}{*}{$\Delta\sigma_{\text{NLO QCD}}$ [pb]} \; 
        {\tt ReneSANCe} & 19.1(1) & 76.5(1) & 76.3(1) & & & \\
        {\tt MCSANC}    & 19.2(1) & 76.6(1) & 76.3(1) & & & \\
        \hline
        \multirow{2}{*}{$\delta_{\text{NLO EW}}$ [\%]} \;
        {\tt ReneSANCe} & -2.05(1) & 0.90(1) & 0.90(1) & 1.78(1) & 3.28(1) & 3.28(1)\\
        {\tt MCSANC}    & -2.05(1) & 0.90(1) & 0.90(1) & 1.77(1) & 3.28(1) & 3.28(1)\\
        \hline
        \multirow{2}{*}{$\delta_{\text{NLO QCD}}$ [\%]} \; 
        {\tt ReneSANCe} & 14.2(1) & 11.2(1) & 11.2(1) & & & \\
        {\tt MCSANC}    & 14.3(1) & 11.2(1) & 11.2(1) & & & \\
        \hline
    \end{tabular}
    \caption{Cross sections and relative corrections for NC and CC DY processes at $\sqrt{s_0} = 1.96$ TeV in $\alpha(0)$-scheme obtained by {\tt ReneSANCe v1.3.0} and {\tt MCSANC v1.21}.}
    \label{tab:Tevatron1_96TeVgf0}
\end{table}

\begin{table}[ht!]
    \centering
    \begin{tabular}{|r|l|l|l|l|l|l|}
        \hline
        $p\bar{p} \to $ & $e^+e^- X$ & $e^+\nu_e X$ & $e^-\bar{\nu}_e X$ & $\mu^+\mu^- X$ & $\mu^+\nu_\mu X$ & $\mu^-\bar{\nu}_\mu X$ \\
        \hline
        \multirow{2}{*}{$\sigma_{\text{LO}}$ [pb]} \; 
        {\tt ReneSANCe} & 132.97(1) & 676.59(1) & 676.61(1) & 132.97(1) & 676.62(1) & 676.62(1)\\
        {\tt MCSANC}    & 132.97(1) & 676.62(1) & 676.63(1) & 132.97(1) & 676.61(1) & 676.61(1)\\
        \hline
        \multirow{2}{*}{$\Delta\sigma_{\gamma q}$ [pb]} \; 
        {\tt ReneSANCe} & -0.064(1) & 0.074(1) & 0.074(1) & -0.065(1) & 0.074(1) & 0.074(1)\\
        {\tt MCSANC}    & -0.065(1) & 0.074(2) & 0.075(2) & -0.065(1) & 0.074(2) & 0.074(2)\\
        \hline
        \multirow{2}{*}{$\Delta\sigma_{\gamma \gamma}$ [pb]} \; 
        {\tt ReneSANCe} & 0.119(1) & $-$ & $-$ & 0.119(1) & $-$ & $-$\\
        {\tt MCSANC}    & 0.119(1) & $-$ & $-$ & 0.119(1) & $-$ & $-$\\
        \hline
        \multirow{2}{*}{$\delta_{\gamma q}$ [\%]} \; 
        {\tt ReneSANCe} & -0.048(1) & 0.011(1) & 0.011(1) & -0.048(1) & 0.011(1) & 0.011(1)\\
        {\tt MCSANC}    & -0.048(1) & 0.011(1) & 0.011(1) & -0.049(1) & 0.011(1) & 0.011(1)\\
        \hline
        \multirow{2}{*}{$\delta_{\gamma \gamma}$ [\%]} \; 
        {\tt ReneSANCe} & 0.089(1) & $-$ & $-$ & 0.089(1) & $-$ & $-$\\
        {\tt MCSANC}    & 0.089(1) & $-$ & $-$ & 0.089(1) & $-$ & $-$\\
        \hline
    \end{tabular}
    \caption{Cross sections and relative corrections for NC and CC DY processes at $\sqrt{s_0} = 1.96$ TeV in $\alpha(0)$-scheme obtained by {\tt ReneSANCe v1.3.0} and {\tt MCSANC v1.21}.}
    \label{tab:Tevatron1_96TeVPhotgf0}
\end{table}

\begin{table}[ht!]
    \centering
    \begin{tabular}{|r|l|l|l|l|l|l|}
        \hline
        $p\bar{p} \to $ & $e^+e^- X$ & $e^+\nu_e X$ & $e^-\bar{\nu}_e X$ & $\mu^+\mu^- X$ & $\mu^+\nu_\mu X$ & $\mu^-\bar{\nu}_\mu X$ \\
        \hline
        \multirow{2}{*}{$\sigma_{\text{LO}}$ [pb]} \; 
        {\tt ReneSANCe} & 144.20(1) & 734.25(2) & 734.24(2) & 144.20(1) & 734.23(2) & 734.21(2)\\
        {\tt MCSANC}    & 144.18(1) & 734.22(1) & 734.21(1) & 144.19(1) & 734.23(1) & 734.23(1)\\
        \hline
        \multirow{2}{*}{$\Delta\sigma_{\text{NLO EW}}$ [pb]} \; 
        {\tt ReneSANCe} & $-11.44(1)$ & $-36.55(2)$ & $-36.56(2)$ & $-5.92(1)$ & $-19.11(1)$ & $-19.09(1)$\\
        {\tt MCSANC}    & $-11.40(2)$ & $-36.53(1)$ & $-36.52(1)$ & $-5.92(1)$ & $-19.08(1)$ & $-19.08(1)$\\
        \hline
        \multirow{2}{*}{$\delta_{\text{NLO EW}}$ [\%]} \; 
        {\tt ReneSANCe} & $-7.94(1)$ & $-4.98(1)$ & $-4.98(1)$ & $-4.10(1)$ & $-2.60(1)$ & $-2.60(1)$\\
        {\tt MCSANC}    & $-7.91(1)$ & $-4.98(1)$ & $-4.97(1)$ & $-4.10(1)$ & $-2.60(1)$ & $-2.60(1)$\\
        \hline
    \end{tabular}
    \caption{Cross sections and relative corrections for NC and CC DY processes at $\sqrt{s_0} = 1.96$ TeV in $G_\mu$-scheme obtained by {\tt ReneSANCe v1.3.0} and {\tt MCSANC v1.21}.}
    \label{tab:Tevatron1_96TeVgf1}
\end{table}

\begin{table}[ht!]
    \centering
    \begin{tabular}{|r|l|l|l|l|l|l|}
        \hline
        $p\bar{p} \to $ & $e^+e^- X$ & $e^+\nu_e X$ & $e^-\bar{\nu}_e X$ & $\mu^+\mu^- X$ & $\mu^+\nu_\mu X$ & $\mu^-\bar{\nu}_\mu X$ \\
        \hline
        \multirow{2}{*}{$\sigma_{\text{LO}}$ [pb]} \; 
        {\tt ReneSANCe} & 142.91(1) & 727.20(1) & 727.20(1) & 142.91(1) & 734.23(2) & 734.21(2)\\
        {\tt MCSANC}    & 142.91(1) & 727.20(1) & 727.20(1) & 142.91(1) & 727.20(1) & 727.20(1)\\
        \hline
        \multirow{2}{*}{$\Delta\sigma_{\gamma q}$ [pb]} \; 
        {\tt ReneSANCe} & $-0.069(1)$ & 0.080(1) & 0.080(1) & $-0.069(1)$ & 0.080(1) & 0.080(1)\\
        {\tt MCSANC}    & $-0.070(9)$ & 0.080(1) & 0.081(1) & $-0.070(9)$ & 0.080(1) & 0.080(1)\\
        \hline
        \multirow{2}{*}{$\Delta\sigma_{\gamma \gamma}$ [pb]} \; 
        {\tt ReneSANCe} & 0.119(1) & $-$ & $-$ & 0.119(1) & $-$ & $-$\\
        {\tt MCSANC}    & 0.119(1) & $-$ & $-$ & 0.119(1) & $-$ & $-$\\
        \hline
        \multirow{2}{*}{$\delta_{\gamma q}$ [\%]} \; 
        {\tt ReneSANCe} & $-0.048(1)$ & 0.011(1) & 0.011(1) & $-0.048(1)$ & 0.011(1) & 0.011(1)\\
        {\tt MCSANC}    & $-0.048(1)$ & 0.011(1) & 0.011(1) & $-0.049(1)$ & 0.011(1) & 0.011(1)\\
        \hline
        \multirow{2}{*}{$\delta_{\gamma \gamma}$ [\%]} \; 
        {\tt ReneSANCe} & 0.083(1) & $-$ & $-$ & 0.083(1) & $-$ & $-$\\
        {\tt MCSANC}    & 0.083(1) & $-$ & $-$ & 0.083(1) & $-$ & $-$\\
        \hline
    \end{tabular}
    \caption{Cross sections and relative corrections for NC and CC DY processes at $\sqrt{s_0} = 1.96$ TeV in $G_\mu$-scheme obtained by {\tt ReneSANCe v1.3.0} and {\tt MCSANC v1.21}.}
    \label{tab:Tevatron1_96TeVPhotgf1}
\end{table}

\clearpage

\clearpage    
    
\clearpage

\clearpage

\clearpage

\clearpage

\clearpage

\clearpage

\clearpage

\section{Summary}
\label{sec:summary}
We presented an extention of the \ReneSANCe~ Monte Carlo generator to the hadron-hadron collisions mode. We performed a tuned comparison of \ReneSANCe~ with the \MCSANC Monte Carlo integrator in NC and CC for DY processes, taking into account realistic requirements of the Tevatron and the LHC. We found perfect  numerical agreement of the prediction for the sets of Z and W observables. Our study includes $\alpha(0)$ and $G_\mu$ EW input schemes and corrections for the contributions for NC and CC DY at the LO EW/QCD and NLO EW/QCD levels, as well as the contributions for photon-induced processes.

\section{Funding}
\label{sec:funding}
The research is supported by grant of the Russian Science Foundation (project No. 22-12-00021). 
 
\section{Acknowledgements}
\label{sec:acknowledgements}
We are grateful to M.~Potapov for the help in preparation of the manuscript.

\bibliographystyle{elsarticle-num}
\bibliography{ReneSANCe_pp}
\end{document}